\documentstyle[11pt,paspconf,epsf]{article}
\def\lesssim{\lower.5ex\hbox{$\; \buildrel < \over \sim \;$}}
\def\gtrsim{\lower.5ex\hbox{$\; \buildrel > \over \sim \;$}} 

\begin{document}

\title{Variability and Unification of Blazars and Gamma Ray
Bursts}

\author{Charles D. Dermer \& James Chiang}
\affil{Naval Research Laboratory, Code 7653, Washington, DC
20375-5352 USA}

\begin{abstract} Most models for blazars and gamma-ray
bursts involve relativistic plasma outflows powered by
accretion processes onto black holes.  The blast wave physics
developed for cosmological models of GRBs is reviewed. Two
points relevant for blazar modeling are made: (1) The
injection of nonthermal relativistic particles in the
comoving frame is simply treated though a process of
energizing the plasma as it sweeps up material from the
surrounding medium. (2) The primary energy source of blazar
radiation derives from the bulk kinetic energy of the
outflowing plasma. Thus deceleration of the plasma blast
wave must be included in blazar flaring calculations, and
this process will introduce temporal and spectral effects in
addition to those produced by acceleration and radiative
cooling.

\end{abstract}


\section{Introduction}

The purpose of this paper is to provide an introductory
review of the blast wave physics developed to model prompt
and afterglow GRB emissions, with an eye toward using these
tools in blazar modeling.  When constructing such models,
an important difference between the
blazar and burst systems that should be kept in mind 
is that blazars involve continuous
accretion processes which expel plasma with initial bulk
Lorentz factors (or baryon-loading factors)
$\Gamma_0 \lesssim 30$.  In contrast, GRBs originate from
one-time catastrophic accretion events which drive plasma
outflows with $\Gamma_0 \sim 300$. Thus a blazar spectrum
could result from a superposition of injection events, whereas
this is less likely to be the case for GRBs unless the 
central source ejects successive shells.

The blazar category refers collectively to several different
classes of sources which are defined according to specific
observational criteria, and includes flat spectrum radio
quasars, highly polarized quasars, optically violently
variable quasars, and BL Lac objects.  Most models
interpret the properties of these classes in view of the
circumstance that our line-of-sight is aligned nearly along the
direction of a jet of relativistic plasma expelled from a
central engine; thus it has proven useful to adopt the
common designation blazar.  

The gamma-ray observations indicate extremely large {\it
apparent} blazar luminosities, attaining values $L_\gamma \sim
10^{48}\Delta \Omega_j$ ergs s$^{-1}$ in some sources (e.g.,
Hartman et al.\ 1997; Mattox et al.\ 1997). The actual
luminosity is reduced from the apparent luminosity due to
jet collimation. If the parent population of blazars are
radio galaxies (e.g., Urry \& Padovani 1995), then jet
beaming factors $\Delta \Omega_j/ 4 \pi \cong
10^{-2}$-$10^{-3}$ are implied by observations, so that
$L\lesssim 10^{46}$-$10^{47}$ ergs s$^{-1}$.  One-day flares
therefore involve energy releases as large as $E\sim
10^{52}$ ergs.

GRBs involve very large, impulsive energy releases. 
In the case of GRB 971214 with a candidate host
galaxy at redshift
$z = 3.42$ (Kulkarni et al. 1998), the apparent
gamma-ray luminosity $L_\gamma \sim 10^{52}$ ergs
s$^{-1}$. The gamma-ray luminous
phase lasted for $\sim 40$ s (Kippen et al.\ 1997). 
Taking into account the afterglow radiation,  this
implies an apparent total energy release of
$E\sim 10^{54}$ ergs. Models of both blazar flares and GRB events therefore
deal with energies per solid angle $\partial E/\partial \Omega 
\sim 10^{53}$ ergs sr$^{-1}$. 
Unlike blazars, whose outflows
are thought to be driven by accretion onto
supermassive black holes, the mechanism producing
GRBs remains controversial.  

Our primary channel of information about blazars and GRBs
derives from the nonthermal radiation produced by the relativistic
plasma expelled from a central engine.  As indicated above,
the total energy release and energy-release time scales
constitute the most important observables, though
determinations of $E$  are uncertain to the extent that the
outflow is collimated.  The next most important quantity to
characterize the explosion physics and nature of the source
is the baryon-loading factor, which is also referred to as
the entropy per baryon.  This quantity represents the mean
energy of a baryon after the fireball has become optically
thin (see, e.g., M\'esz\'aros \& Rees 1993; Piran 1994),
 and is essentially equivalent to 
the initial Lorentz
factor $\Gamma_0$ of the bulk outflow.  

In what follows, we review the physics that has been
developed to analyze GRB observations, and indicate how
these results can be applied to blazar research and  other
systems which produce relativistic plasma outflows.  By
characterizing the properties of these outflows, we hope to
determine the nature of the sources which expel highly
relativistic plasma.

\section{Fireball and Blast Wave Physics}

If a mass $M_{\rm baryon}$ of baryons is mixed in the
explosion energy, then the resulting fireball will expand
and transfer most of its energy to baryons, ultimately producing a
relativistic outflow moving with Lorentz factor
\begin{equation}
\Gamma_0 \simeq { E_{\rm tot}\over M_{\rm baryon} c^2}.
\end{equation}

Observations of superluminal motion in radio-loud AGNs
reveal parsec-scale relativistic plasma outflows moving with
a wide range of $\Gamma_0$, though rarely exceeding $\Gamma_0
\sim 20$-$ 30$ (e.g., Vermeulen \& Cohen 1994). An outstanding question 
in black hole
research is the nature of the processes which expel
relativistic plasma.  Many mechanisms have been suggested,
including the Blandford-Znajek (1977) process for extracting black
hole rotational energy through magnetic couplings between
the accretion disk and black hole.  The large radiant luminosities
of  accreting black holes have also been suggested to drive hot
plasma to relativistic energies through the Compton rocket
effect (e.g., O'Dell 1981; Phinney 1982), or to slow very
 high energy particles through Compton drag (Melia \& K\"onigl 1989). 
 Stochastic acceration with accompanying
diffusive escape of high-energy protons and electrons has also been
proposed (Dermer, Miller, \& Li 1996) as a source of outflows 
from accreting black holes.

As already mentioned, the nature of the central engine which drives
the relativistic outflows from GRB sources is unclear.
It is interesting to note that if we assume the emission is
uncollimated in GRB 971214 and  that
$\Gamma_0 = 300 \Gamma_{300}$ with $\Gamma_{300}\sim 1$,
then only a Jupiter's mass of baryons is mixed into an
energy release $\gtrsim 100$ times the entire energy that
the Sun will radiate over its lifetime.  This circumstance
requires an extraordinarily baryon-free environment. The
explanation of these incredibly energetic outflows is one of
the outstanding contermporary problems in astronomy. Widely discussed 
GRB triggers include coalescence of compact objects (e.g, M\'esz\'aros,
 \& Rees 1992), the collapse
of massive stars (Paczy\'nski 1998), and failed supernovae or
collapsars (Woosley 1993).  

\subsection{Special Relativity} Special relativity plays a
key role in spectral calculations of  relativistic plasma
outflow. The most important effects include the relativistic boosting
of the radiation, which is primarily governed by the value
of the Doppler factor ${\cal D} =
[\Gamma(1-B\mu)]^{-1}$, where  $B =
\sqrt{1-\Gamma^{-2}}$ and $\arccos\mu$ is the angle
between the observer and the direction of motion of the
radiating element. The Doppler boosting means that most
of the observed radiation is emitted from angles
\begin{equation}
\theta \lesssim 1/\Gamma\; .
\end {equation} A second important effect is that during the time
interval $\Delta t$ over which an observation takes
place, the plasmoid traverses a distance
\begin{equation}
\Delta x \cong \Gamma^2 c \Delta t/(1+z)\; .
\end {equation} A third effect is the strong reduction of the
comoving luminosity $L^\prime$ by a factor
$\sim \Gamma^4$ in comparison with the luminosity that would
be inferred if the radiating plasma is assumed to be at rest
with respect to the observer.  If
$J(\epsilon^\prime)$ (ergs s$^{-1}$
$\epsilon^{\prime~-1}$) is the spectral emissivity at
comoving dimensionless photon energy $\epsilon^\prime =
h\nu^\prime/(m_ec^2)$, then 
\begin{equation}
\nu L_\nu = {\cal D}^4\epsilon^\prime J(\epsilon^\prime)
\sim \Gamma^4 L^\prime \; .
\end {equation} These effects are well-known when $\Gamma$
is constant (see, e.g., Begelman et al. 1984, Dermer et al.
1997). Variations of $\Gamma$ with distance (or time) 
will complicate these relations. Much of the new physics
being studied as a result of GRB observations has to do
with observable consequences resulting from deceleration 
of the relativistic outflows.

The effect described by eq.(4) resolves 
the $\gamma\gamma$ transparency problem in blazars and GRBs. 
If the emitting region 
were stationary with a size $R\sim
c\Delta t_{\rm var}/(1+z)$ implied by the variability time
scale $\Delta t_{\rm var}$, then $\gamma$-rays could not 
escape due to attenuation by the process 
$\gamma+\gamma^\prime\rightarrow e^+ + e^-$ (Baring \& Harding 1997;
Maraschi et al. 1992). Given the
rapid variability of the radiation, either
beaming of the radiation or bulk relativistic outflows are
therefore required to explain the detection of gamma rays.
Depending on the specific GRB and the observing time, the emitting regions
 of GRBs must be moving  with $\Gamma_0\gtrsim 30$-100 in order to be
transparent to gamma rays.

\subsection{Deceleration Radius and Time Scale}

As relativistic plasma expands into the surrounding medium,
its entrained magnetic field sweeps up and captures electrically-charged 
particles. These particles enter
the comoving fluid frame with Lorentz factor $\Gamma$ to
provide a source of free energy which becomes available to
be radiated.  The strength of the plasma magnetic field
necessary to capture particles is determined by the condition
that the Larmor radius is less than the width of the
plasmoid or blast wave. Here we distinguish the use of the
terms ``plasmoid" and ``blast wave."   Plasmoid refers to
any coherent plasma outflow that can be treated as a
hydrodynamical fluid.  The plasmoid cross-sectional area which is effective
at sweeping up particles is $A(x) \propto
f(x)\delta \Omega$, where $f(x)$ can be specified generally
and $x$ is a spatial coordinate. By contrast, a blast wave
is a type of plasmoid which expands according to the rule
$A(x)\propto x^2$. Henceforth we consider blast waves,
keeping in mind that the results can be straightforwardly
extended to the more general plasmoid geometry.

The total mass which is swept-up by the plasma as it travels
a distance $\delta x$ is simply $m_p n(x)A(x)\delta x$,
where $n(x)$ is the density of the particles in the
surrounding medium, which we suppose to be composed
 of hydrogen. Each proton and electron captured
from the surrounding medium carries a relativistic energy
$\Gamma m_pc^2$ into the comoving fluid frame.
 The blast wave no longer coasts but
decelerates when the incoming relativistic inertia equals
the rest mass energy of the ejecta, that is, when $V_d m_p
\Gamma_0 c^2 = E/(\Gamma_0 c^2)$, where the deceleration
volume $V_d = 4\pi x_d^3/3$.  Hence the deceleration radius
is given by
\begin{equation} x_d = ({3 E\over 4\pi n_0\Gamma_0^2
m_p})^{1/3}\cong  2.6\times 10^{16} ({E_{54}\over n_2
\Gamma_{300}^2})^{1/3}\;{\rm cm}\;
\end{equation}
(Rees \& M\'esz\'aros 1992), where $n_0 = 10^2 n_2$ cm$^{-3}$ is the
density at $x_d$, the explosion energy $E = 10^{54}E_{54}$ ergs,
  and we assumes a uniform density $n(x) =
n_0$.

The strongly Doppler-beamed emission from a relativistically
expanding shell means that only that portion of the blast
wave found within an angle
$\sim 1/\Gamma_0$ to the line-of-sight produces much of the
prompt GRB emission. In the frame of the observer, the
emitting region is chasing its photons and keeping close
behind them, at least when
$\Gamma/\Gamma_0\sim 1$.   The $(1-B\mu)$ factor in the Doppler effect
means that the deceleration time scale $t_d$ of the
blast-wave, as measured by an observer, is given by
\begin{equation} t_d = {(1+z)x_d\over \Gamma_0^2 c}\;\cong
9.6 (1+z)({E_{54}\over n_2\Gamma_{300}^8})^{1/3}\;{\rm s} .
\end{equation}

\subsection{Radiative Regimes}

The equation of motion of the blast wave depends on the
amount of internal energy it radiates.  The adiabatic regime
applies to the case where very little of its internal energy
is radiated, and the radiative regime
applies when essentially all of its internal energy is
radiated on a time scale short compared with the deceleration time
scale as measured in the comoving frame.  In
the former case, we can write the equation of momentum
conservation as
\begin{equation}
B\Gamma M_{\rm baryon} + (B\Gamma ) \cdot
\Gamma\cdot ({4\pi\over 3} m_p n_0 x^3) = B_0\Gamma_0
M_{\rm baryon}\; .
\end{equation} When $x \gg x_d$,
\begin{equation}
\Gamma_{\rm adi} \propto x^{-3/2}\; \;
\end{equation}
(Blandford \& McKee 1976). 
In the latter (radiative) case, the momentum conservation
equation is the same as given in eq.\ (7), except that the
weighting factor $\Gamma$ in the second term on the left-hand-side
 is replaced by unity, because all the internal
energy is promptly radiated away.  Thus we find that when $x
\gg x_d$,
\begin{equation}
\Gamma_{\rm rad} \propto x^{-3}\; \;.
\end{equation} In general,
\begin{equation}
\Gamma_{\rm rad} \propto x^{-g}\; ,\; {\rm with~} 3/2 < g <
3\; .
\end{equation}
The range of $g$ specified in eq.(10) 
applies to the case of a uniform density surrounding
medium.

\subsection{Properties of the Relativistic Plasma} 
The baryons initially mixed in the explosion
adiabatically expand until the initial explosion
energy is transformed into directed plasma kinetic
energy.  In the comoving fluid frame, the baryons
become nonrelativistic. Particles which are swept-up
from the surrounding medium with Lorentz factor
$\Gamma$ in the comoving frame  provide a source of
free energy which is derived from the bulk kinetic
energy. This causes the blast wave to decelerate,
and the Doppler boosting becomes weaker.  Both
nonthermal electrons and protons are swept up, but
because of the greater mass of the protons, most of
the free energy is initially in the form of nonthermal
protons. Other than to simply heat the thermal plasma
or escape, the only known way to radiatively
dissipate this energy efficiently is by transfering
a large fraction of the nonthermal proton energy to
nonthermal electrons. 

A nonthermal proton could in principle transfer
nearly all its energy to a very energetic electron.  This
maximum defines a minimum Lorentz factor $\gamma_{\rm
e,min}$ carried by a swept-up electron, noting that
Fermi processes in the blast wave could further
accelerate the nonthermal electrons to
$\gamma \gg \gamma_{\rm e,min}$. Thus 
\begin{equation}
\gamma_{\rm e,min} = \xi_e (m_p/m_e)\Gamma \;,
\end{equation} 
where the electron equipartition factor
$\xi_e\lesssim 1$.   

A central uncertainty in blast wave physics is the
magnetic field strength in the comoving fluid frame.
A convenient way to characterize the blast wave
magnetic field $H$ is through the
prescription
\begin{equation} {H^2\over 8\pi} \cong (4\Gamma
n_0)\;(\Gamma m_p c^2) \xi_H \; ,
\end{equation} where the right hand side represents
the energy density of the swept-up electron-proton
fluid downstream of the blast wave shock. Note that
the particle density is enhanced by Lorentz
contraction, and that the compression ratio factor $r$ is
normalized to the factor of 4 which represents an
important limit in nonrelativistic shock theory (a
more realistic value of $r$ would be 7, given that
the downstream particles constitute a relativistic
fluid).  The term
$\xi_H$ is the magnetic equipartion parameter.  Thus 
\begin{equation} H \cong 120 \;\Gamma_{300} n_0^{1/2}
\xi_H^{1/2}\;{\rm Gauss}\; .
\end{equation}

The relativistic electrons in the blast-wave
plasma will radiate nonthermal synchrotron
radiation with frequencies $\nu_p[{\rm Hz}] \gtrsim
3\times 10^6 H\gamma_{\rm e,min}^2$ $ \propto
\Gamma_{300}^3 \xi_H^{1/2}\xi_e^2$ in the blast-wave frame. If the
index $p$ of the electron number spectrum is steeper
than 3, this frequency represents the peak of the
measured $\nu F_\nu$ spectral energy distribution
produced by nonthermal electron synchrotron. Other
spectral components, in particular the
synchrotron self-Compton (SSC) and the nonthermal
proton synchrotron emissions, will also be radiated,
but are usually found (Chiang
\& Dermer 1998; Panaitescu
\& M\'esz\'aros 1998; B\"ottcher \& Dermer 1998) to be
energetically less important than the nonthermal
electron synchrotron luminosity during the prompt
gamma-ray luminous phase of a GRB.

\subsection{Sweeping Energization}

The conversion of the bulk kinetic energy of the
plasma outflow to nonthermal particle and magnetic-field 
energy in the comoving frame involves very
complicated plasma physics and shock acceleration
processes. If this energy is extracted through the
process of sweeping-up material from the surrounding
medium, the detailed acceleration physics can
fortunately be avoided by simply assuming that some
fraction of the initial swept-up energy is
transformed into a distribution of nonthermal
electrons with a minimum energy given by eq.(11).  

The correct description of the plasmoid dynamics must
take into account the changing internal energy
content and inertia of the plasmoid as it sweeps up
matter from the external medium.  The momentum
distribution of the swept-up energetic particles
evolves through subsequent acceleration, radiative losses,
and particle escape.  This causes the inertia of the
plasmoid to change, thereby affecting its subsequent
dynamics.  The important point to note is that {\it
the energy of the nonthermal particle distribution
function in the comoving frame derives ultimately
from the directed bulk kinetic energy of the plasma
outflow; thus models for GRBs and blazar flares must
include effects of plasmoid deceleration. } 

An equation which explicitly couples the plasmoid
dynamics with the comoving particle distribution
function was derived from momentum conservation by
Dermer
\& Chiang (1998), and is given by
\begin{equation} -{[\partial P(x)/\partial x]\over
P(x) } = { n_{\rm ext}(x) A(x) \Gamma(x)(1+m_e/m_p) 
\over N_{\rm th}(1+a_{\rm th}) +
\int_0^\infty dp\;
\gamma
\; [N_{\rm pr}(p;x) + a_{\rm nt} N_{\rm e}(p;x)]}\;.
\end{equation} Here $p=\beta\gamma$ is the
dimensionless momenta of the protons or electrons
whose momentum distribution functions are denoted by
the subscript ``pr" or ``e", respectively, and 
$P = B\Gamma$ is the dimensionless momentum of the
plasmoid. In eq.(14), only protons and electrons are
considered, though the result is easily generalized for
heavier ions and pairs. This expression represents the
inverse of the characteristic length scale over which the plasmoid
decelerates. The rate of spatial deceleration is
impeded by the relativistic inertia of the plasmoid,
which is given by the two terms in the denominator on
the right-hand-side.  The first term represents the
number of thermal protons multiplied by a weighting
factor which corrects for thermal electrons or
pairs.  If only electrons are present,
$a_{\rm th} = m_e/m_p$.  The second term in the
denominator in the form of an integral  represents
the relativistic inertia given through the nonthermal proton and
electron distribution functions. Again, if no pairs
are present,
$a_{\rm nt} = m_e/m_p$.  The numerator on the
right-hand-side of eq.(14) is proportional to the
spatial momentum impulse applied to the plasmoid
through the process of sweeping-up particles from the
external medium.

Eq.(14) is amenable to a variety of analytic
treatments, and can easily be shown to yield the
adiabatic and radiative limits previously considered.
An analytic solution is possible for a plasmoid with
a constant entrained magnetic field which decelerates
by sweeping up material from a smooth external medium
with density
$n_{\rm ext}(x) \propto x^{-\eta}$ (Dermer \& Chiang
1998).  There we showed that GRB afterglows
exhibit two temporal and spectral regimes, depending
on whether synchrotron cooling plays a negligible or
major role in the cooling of the bulk of the
electrons that produce radiation at a given energy.
If the flux density is written in the form
$S(\epsilon,t) = \epsilon\dot N(\epsilon,t) =
Kt^{-\chi}\epsilon^{-\alpha}$, then in the uncooled
regime $\alpha_{\rm uncooled} = (s-1)/2$, where $s$
in the injection index of electrons, and 
\begin{equation}
\chi_{\rm uncooled} = {g(3+s)-2(j+1-\eta)\over
2(2g+1)}\rightarrow {g(s+3)-6\over 2(2g+1)}\;.
\end{equation} In the cooled regime, $\alpha_{\rm
cooled}= s/2$ and 
\begin{equation}\chi_{\rm cooled} =
{g(s+6)+2-2(j+1-\eta)\over 2(2g+1)}\rightarrow
{g(s+6)-4\over 2(2g+1)}\;.
\end{equation} The transition
from the uncooled to the cooled regime occurs at the observer time
\begin{equation}
t_{\rm tr} \simeq {5(1+z)^{0.38}\over
H_0^{1.85}\epsilon^{0.62}}\;\;{\rm sec}
\end{equation}
(see eqs.\ 52-54 in
Dermer \& Chiang 1998), where $\epsilon$ is the observed dimensionless
photon energy and $H_0$ is the magnetic field
strength in Gauss, assumed constant in time. The expressions 
on the rhs of eqs.(15) and (16) apply to a blast-wave geometry (i.e.,
$A(x)\propto x^j$ with $j=2$) and a uniform density ($\eta =
0$) medium. 

It is a simple matter to show that if the X-ray afterglows are
observed in the cooled regime, as is likely,
 then injection indices $s \cong 2$ and radiative indices in the range
$3/2 \lesssim g \lesssim 3$ imply $1.0\lesssim \chi \lesssim 10/7 =
1.43$. This reproduces the range of temporal indices observed in
GRB X-ray afterglows, which are generally well fit by $\chi$ in the range $1.1\lesssim \chi
\lesssim 1.5$  (e.g, Costa et al.\ 1997;  Feroci et
al.\ 1998; Piro et al.\ 1998). This analytic treatment (see also Sari,
Piran, \& Narayan 1998; Wijers et al.\ 1997) makes several
predictions, most importantly that during the transition from the
uncooled to the cooled regime, the photon spectral index changes by
1/2 unit and the temporal index changes in value according to eqs.(15)
and (16). 

Eq.(14) can also be used to make estimates for the temporal and
spectral behavior associated with sweeping energization and blast
wave deceleration by other radiation processes.  For example,
very high energy protons will produce nonthermal synchrotron
radiation (Vietri 1997).  Because the protons radiate very
inefficiently, they are, in general, not strongly cooled, in which
case the emission from this process can be shown (B\"ottcher
\& Dermer 1998) to decay more slowly than electron
synchrotron or SSC radiation. Temporal decay indices for proton 
synchrotron radiation from uncooled protons in a constant 
magnetic field are given by eq.(15); the indices in the uncooled
synchrotron case for a magnetic 
field which varies according to eq.(13), with $\xi_H$ constant,
 are given by
(B\"ottcher \& Dermer 1998).

\subsection{Blast Wave Deceleration}

In blazar physics, nonthermal electrons are
normally injected either impulsively or over some fixed
interval of time in the comoving frame.  The effects
of particle acceleration or cooling processes are
then calculated and used to model the multiwavelength
spectral energy distributions, or to characterize observations of 
spectral variability. Blast wave physics shows how the
injection problem can be solved, namely by extracting
energy from the bulk relativistic motion of the
plasmoid and transforming it into nonthermal
particle energy, making sure that momentum
conservation is satisified. The major uncertainty
in following this prescription is to specify the
density distribution of the surrounding medium which
is intercepted and swept-up by the plasmoid. A more comprehensive
calculation, beyond the one-component description discussed here, should 
also take into account radiation from the reverse shock 
(see, e.g., Panaitescu, \& M\'esz\'aros 1998)
and light-travel time effects from injection into different
portions of a plasmoid
or blast wave with finite extent.  

The discussion in the previous section illustrates the importance
of treating particle injection and blast wave
deceleration on equal footing. For example, a decaying time
profile cannot be simply interpreted as a
decrease in the number of radiating electrons 
through either cooling or losses. Because the injection of
nonthermal particles is accompanied by plasmoid
deceleration,  it is equally possible that the number of
nonthermal particles increases, yet a decreasing
flux is observed due to the slowing-down of the
blast wave.  In fact, this is the precisely the behavior
that is evidently observed in GRBs at times $t > t_d$.

The interplay and importance of these effects are shown
by models of X-ray
observations of blazar sources such as Mrk 421 (Takahashi et
al.\ 1996),  OJ 287 (Idesawa et al.\ 1997), and PKS
2155-304 (Sembay et al.\ 1993).  The X-ray data
follow well-defined trajectories in the spectral
index/flux plane.  This behavior can be explained
by synchrotron losses of nonthermal electrons which are
injected into the comoving frame
over a finite length of time (Dermer 1998; Kirk,
Rieger, \& Mastichiadis 1998).  This behavior can also,
however, be a consequence of the effects of plasmoid
deceleration (Chiang 1998; Dermer, Li, \& Chiang
1998). A related instance is the interpretations of energy-
dependent lags from blazar flares.  For example, the UV radiation lagged
the X-ray emission in a flare from the BL Lac object PKS 2155-304 (Urry
et al.\ 1994). This behavior could result from blast wave deceleration
rather than from synchrotron losses. 

\section{Unification of Blazars and GRBs}

Here we speak of unification in terms of the
underlying physical processes which produce the
radiation from the relativistic plasma outflows in
blazars and GRBs. The differences between the engines
driving blazar and GRB outflows are obviously
profound, but can be better investigated once the
origin of these radiant emissions is established.
Because blast wave physics has been used to
successfully explain GRB afterglows, we argue
that these tools should be equally applied to blazars.

\setcounter{figure}{0}
\begin{figure}
\epsfysize=13cm
\epsffile[60 180 600 700]{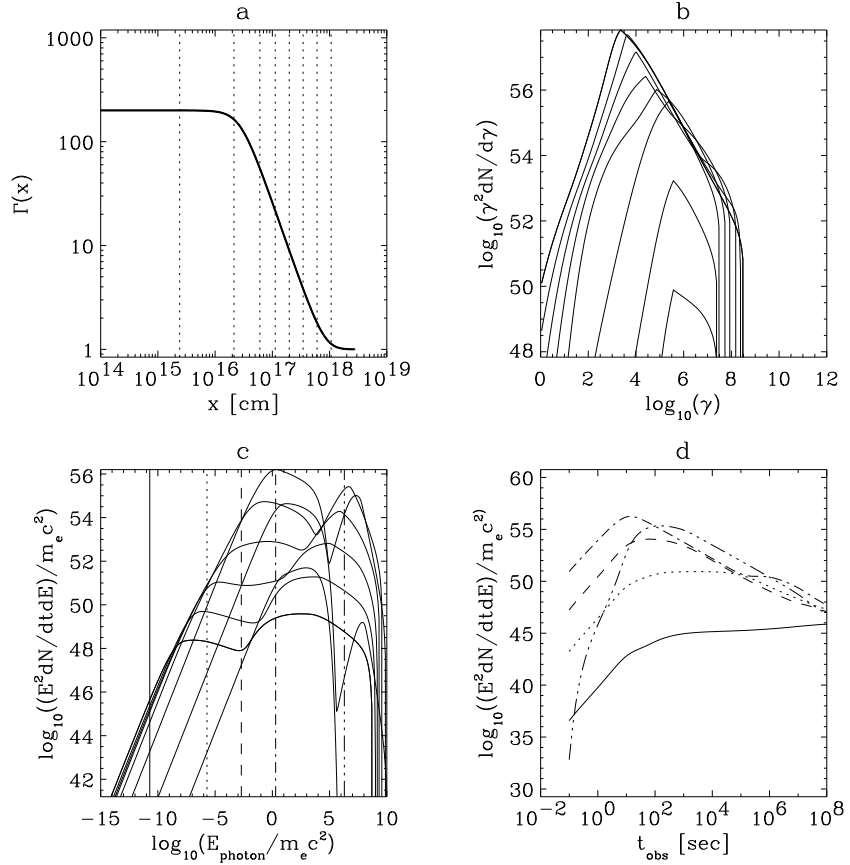}
\caption[]{Calculations of blast wave deceleration
and radiation for a fireball with energy $E = 10^{54}$ ergs and 
$\Gamma_0$ = 200, assuming that 10\% of the swept-up
nonthermal particle energy goes into nonthermal
electrons. The dotted lines in panel a mark the
Lorentz factors at observer times of 1 s, 10 s, 100 s, etc.
Nonthermal electron spectra
(panel b) and $\nu L_\nu$ photon spectra (panel c) are
calculated in panels b and c, respectively.
 The respective Lorentz factors and times
($\Gamma_0$,t) for the electron and photon spectra
are (200.0, 0.1 s), (199.8, 1.4 s), (122, 19.0 s),
(36.9, 261.6 s), (13.2, 3600 s), (4.9, $5.0\times
10^{4}$ s), (2.0, $6.8\times 10^{5}$ s), and (1.14,
$9.4\times 10^{6}$ s). Panel d shows light curves at
radio, optical, X-ray, MeV, and TeV energies.  The
shortest flaring time scale for this calculation is
about 10 seconds at gamma-ray energies.
 }
\end{figure}

Figs.\ 1 and 2 show calculations of the
synchrotron and SSC radiation produced by
uncollimated relativistic outflows which sweep up
material from the surrounding medium, using the code
developed by and described in Chiang \& Dermer (1998). These
calculations show spectra produced by uncollimated fireballs with
total energy $E = 10^{54}$ ergs which are surrounded by a
medium with uniform density $n_0 = 100$ cm$^{-3}$.  If
the fireball is collimated into
$\Delta\Omega_j$ sr, as might be the case for a
blazar, then these calculations would
equally represent the case where an observer
was looking down a jet of material from a fireball
where $10^{54}\Delta\Omega_j/(4\pi)$ ergs are
injected, provided that the plasmoid area increased
$\propto x^2$ and that the jet opening angle $\gg \Gamma_0^{-1}$.

\begin{figure}
\epsfysize=13cm
\epsffile[60 180 600 700]{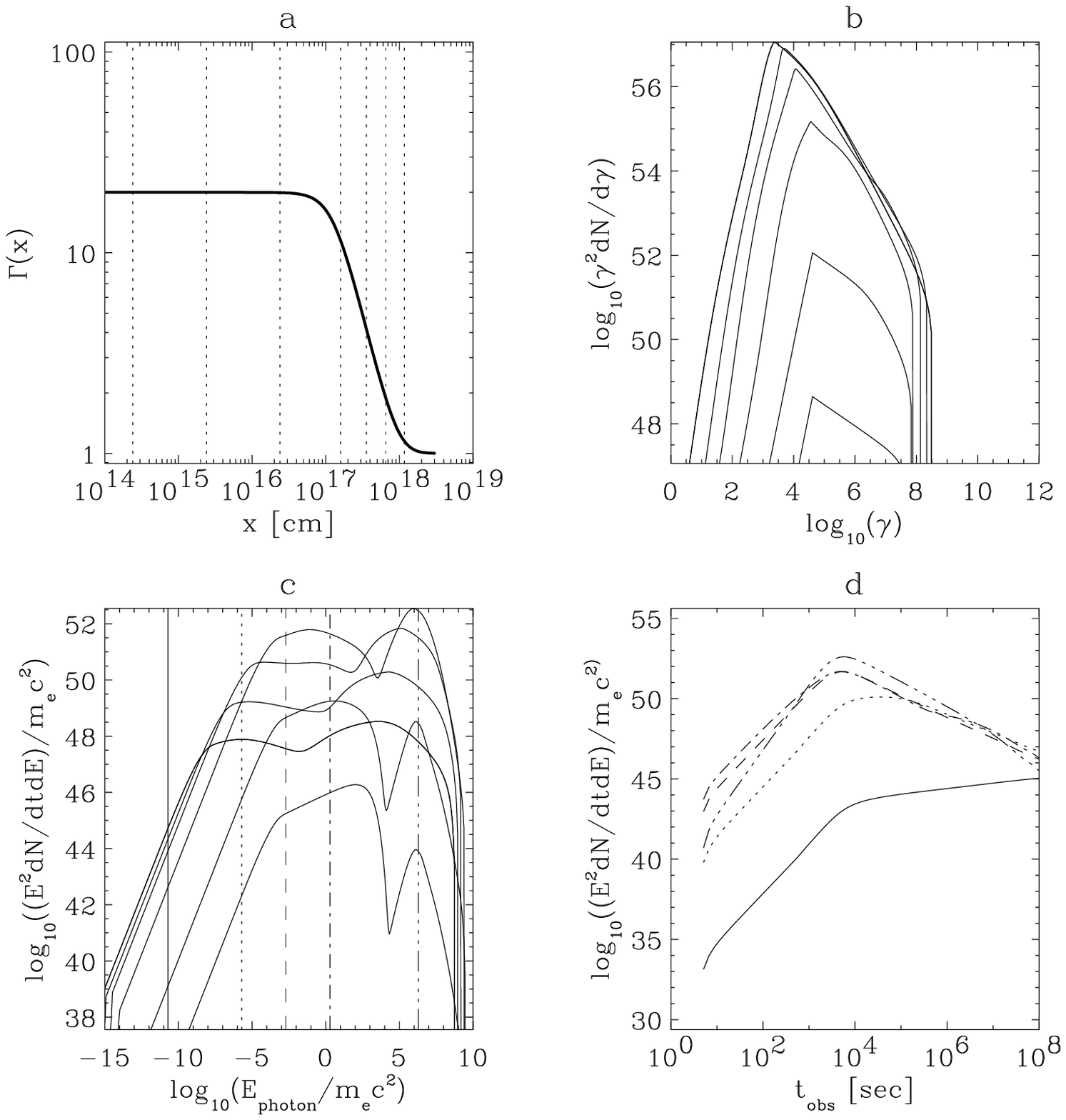}
\caption[]{ Calculations of blast wave deceleration
and radiation for a fireball with $E = 10^{54}$ ergs and
$\Gamma_0$ = 20, assuming that 1\% of the swept-up
nonthermal particle energy goes into nonthermal
electrons. The dotted lines in panel a mark the
Lorentz factors at observer times of 10 s, 100 s, 1000 s, etc.
Nonthermal electron spectra
(panel b) and $\nu L_\nu$ photon spectra (panel c) are
calculated in panels b and c, respectively.
The respective
Lorentz factors and times ($\Gamma_0$,t) for the
electron and photon spectra are (20.0, 19.0 s),
(20.0, 262 s), (17.8, 3600 s), (5.5, $5.0\times 10^{4}$ s),
(2.1, $6.8\times 10^{5}$ s), and (1.16, $9.4\times
10^{6}$ s). Panel d shows light curves at radio,
optical, X-ray, MeV, and TeV energies. The shortest
flaring time scale for this calculation is about an
hour at X-ray, gamma-ray and TeV energies. }
\end{figure}

Only two parameters differ between Figs.\ 1
and 2, namely $\Gamma_0$ and the fraction $\xi$ of
swept-up energy that is promptly channeled into nonthermal
electrons.  In Fig.\ 1, $\Gamma_0 = 200$ and $\xi =
0.1$, whereas in Fig.\ 2, $\Gamma_0 = 20$ and $\xi =
0.01$.  In both calculations, $\xi_e = 1$ and
$\xi_H = 10^{-5}$ (see Chiang \& Dermer 1998 for reasons to
assign this value in order to match GRB observations), 
the electron injection index $s =
3$, and $\gamma$ ranges from $\gamma_{\rm e,min}$
given by eq.(11) to $\gamma_{\rm e,max}\cong
10^6/\sqrt{H({\rm G})}$.  This latter value is a
theoretical upper limit determined by equating the
 time scale for an electron to execute a Larmor orbit
with the synchrotron energy
loss time scale.

Fig.\ 1 provides a reasonable representation of the
typical spectral and temporal behavior of a
``classical" GRB.  The prompt emission, lasting
for $\sim 10$ s (compare eq.[6]), is dominated
by power emitted at gamma-ray energies.  At later
times, power-law X-ray and optical afterglows
are found.  The peak of the $\nu F_\nu$ spectrum
moves to lower energies with time due to blast
wave deceleration and synchrotron cooling. Two
peaks in the spectral energy distribution are
formed.  The higher energy peak is formed from the
SSC process, and produces late time GeV emission, as
has been observed from GRBs (e.g., Hurley 1994).

As in the case of Fig.\ 1, Fig.\ 2 also shows a twin-peaked
spectral energy distribution, which corresponds to the 
generic shape of the  multiwavelength blazar spectra.
 The peak of the synchrotron
component is now at X-ray energies rather than gamma-ray
energies. The SSC component peaks at $\sim 1$ GeV - TeV
energies.  For this choice of parameters, the shortest
flaring time scale is
$\approx 1$ hour due to the $\Gamma_0^{-8/3}$
dependence of $t_d$ given by eq.(6).  This overall
behavior is in rough agreement with the flaring
characteristics of BL Lac objects such as Mrk 421
and Mrk 501 (Macomb et al.\ 1995; Buckley et al.\
1996; Catanese et al.\ 1997). A better comparison with the 
spectral energy distribution of X-ray selected BL Lacs might
require that the upper energy of the
electrons is less than the maximum value given by $\gamma_{\rm e,max}$.

\section {Summary and Conclusions}

New developments in our understanding of GRB
sources appear to have much relevance for blazar
modeling.  Here we have tried to explain the basics of 
blast-wave physics, and to show how this approach can be 
straightforwardly applied to models of variability behavior 
in blazars. As illustrated in Figs. 1 and 2, models employing
blast-wave physics qualitatively describe the
overall spectral and temporal characteristics of GRBs and some 
X-ray selected BL Lac objects for which multiwavelength 
campaigns have been conducted.  As higher quality X-ray and TeV
observations become available, this model will provide specific 
predictions for correlated spectral and temporal behavior.
Before applying our model to flat spectrum radio quasars, however,
extensions of this work to deal with external radiation fields
 must be included.

The central point made here is that blazar flares are a result of
the conversion of directed plasma energy into internal nonthermal
particle energy. Energization of the nonthermal protons 
and electrons comes at the expense of the directed kinetic
 energy of the relativistic 
outflow.  Associated blast-wave deceleration must therefore
be treated when modeling blazar flares. Blazars and GRBs display 
overall similarities in terms of their spectral energy distributions.
 To first order, blazars  differ from GRBs by ejecting plasma
 with $\sim $ an order-of-magnitude 
less entropy per baryon. Why this should be so is unknown, and
 represents a significant clue and 
a strong constraint on any models for the underlying engines.

\baselineskip = 12pt

\section{Acknowledgments}
Useful discussions with M. B\"ottcher are acknowledged.
The work of JC was performed while he held a National Research
Council - Naval Research Laboratory Associateship.
The work of CD is supported by the Office of Naval Research.

\baselineskip = 12pt


\end{document}